\documentclass[12pt]{iopart}

\expandafter\let\csname equation*\endcsname\relax
\expandafter\let\csname endequation*\endcsname\relax

\usepackage{graphicx}%
\usepackage{multirow}%
\usepackage{amsmath,amssymb,amsfonts, amsthm}%
\usepackage{subcaption}%
\usepackage{mathrsfs}%
\usepackage[title]{appendix}%
\usepackage{xcolor,xurl,hyperref}%
\usepackage{textcomp}%
\usepackage{manyfoot}%
\usepackage{booktabs,tabularx}
\usepackage{algorithm}%
\usepackage{algorithmicx}%
\usepackage{algpseudocode}%
\usepackage{listings}%
\usepackage{verbatim}
\usepackage{mathtools}
\usepackage{natbib}

\theoremstyle{plain}

\theoremstyle{definition}

\theoremstyle{remark}

\begin{document}

\title[]{A Generalized Gompertz Law for the Rise and Fall of Empires}

\author{Sabin Roman$^1$, Vaishnavi Jayakumar$^2$ and Karoline Wiesner$^2$}

\address{$^1$Department of Knowledge Technologies, Jo\v{z}ef Stefan Institute, Slovenia}
\address{$^2$Institute for Physics and Astronomy, University of Potsdam, Germany}

\ead{sabin.roman@ijs.si}

\begin{abstract}
Human societies tend to evolve toward increasing scale and complexity, often followed by stagnation and decline. Historical empires provide a prominent example of this trajectory; however, existing quantitative models capture expansion and rarely account for decline within a unified framework. Here, we introduce a minimal dynamical model that describes both growth and collapse in a single formalism. The model yields a closed-form trajectory that generalizes the Gompertz law and admits equivalent maximum-entropy and optimal-control formulations. We show that this functional form captures the rise and fall of fourteen diverse empires across widely varying time scales. Despite substantial historical heterogeneity, all cases follow a common asymmetric rise–peak–decline pattern governed by a small number of interpretable parameters. We identify the mechanism driving this behavior as the chronophage: an exponentially accumulating burden of administrative and coordination costs that offsets the gains of expansion. These findings uncover a general quantitative principle, yielding new insights into how increasing complexity constrains large-scale social systems.
\end{abstract}

%
% Uncomment for keywords
%\vspace{2pc}
%\noindent{\it Keywords}: XXXXXX, YYYYYYYY, ZZZZZZZZZ
%
% Uncomment for Submitted to journal title message
%\submitto{\JPA}
%
% Uncomment if a separate title page is required
%\maketitle
% 
% For two-column output uncomment the next line and choose [10pt] rather than [12pt] in the \documentclass declaration
%\ioptwocol
%

\section{Introduction}

Complex systems across the natural and social sciences often exhibit characteristic trajectories of growth, saturation, and eventual decline. Such dynamics are observed in domains as diverse as organismal ageing, tumour progression, and population growth, where evolution of different kinds proceeds through asymmetric rise–peak–decline patterns rather than indefinite expansion; a broad class of these processes is well described by Gompertz-type dynamics \citep{Kirkwood2015, small2, tumor1, diseasecovid1, mammal1}. 

\begin{figure}[t]
\centering
\includegraphics[width=\linewidth]{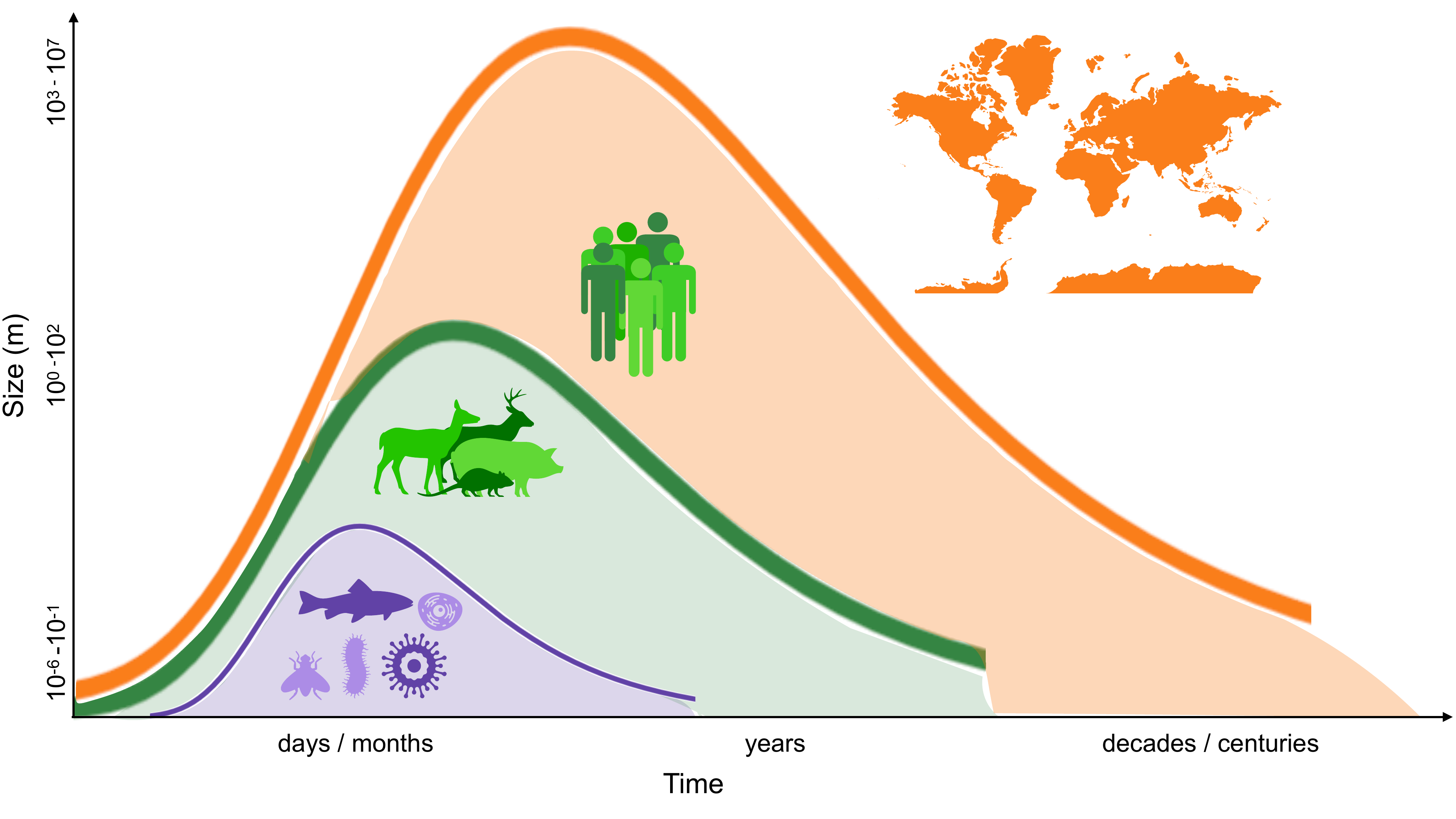}
\caption{
\textbf{Illustration of Gompertz-type growth and decline dynamic of  living systems at multiple scales.} Gompertz-like functions govern a variety of biological and social mechanisms at different length and time scales, ranging from multicellular life to epidemic outbreaks. The life cycle of various empires also follows a similar pattern, as shown by the fit of Eq.~\eqref{eq:solution} to historical data.
}
\label{fig:scales}
\end{figure}

Historical empires provide a particularly rich and well-documented class of complex systems in which these dynamics unfold at large spatial and temporal scales. A long-standing question across history and the social sciences is: Why such systems, after sustained periods of growth, ultimately decline and collapse? Existing explanations have emphasized a wide range of mechanisms, including environmental stress, external shocks, and internal conflict \citep{butzer2012collapse,cumming2017unifying,roman2023collapse}. A prominent theoretical perspective, articulated by Joseph Tainter, attributes collapse to diminishing marginal returns on increasing societal complexity, whereby the costs of maintaining large-scale organization eventually outweigh its benefits \citep{Tainter1, roman2025long}. While these perspectives provide important insights, they have not been integrated into a concise quantitative framework capable of jointly describing growth and decline in a low-dimensional setting.

Previous quantitative approaches have successfully captured the expansion phase of empires. In particular, models of logistic growth reproduce the territorial evolution of diverse historical cases using a small number of parameters, suggesting the presence of coarse-grained regularities despite substantial historical heterogeneity \citep{marchetti2012quantitative}. However, such models are intrinsically limited to monotonic growth and do not account for the subsequent loss of territory and eventual collapse. As a result, the full life cycle of large-scale social systems remains without a unified, minimal dynamical description.

Here, we introduce a general dynamical framework that captures both growth and decline within a single, parameter-sparse formalism. The model yields a closed-form trajectory describing rise, peak, and decline, which generalizes the Gompertz law and can be derived independently from maximum-entropy and optimal-control principles. Within this framework, decline emerges endogenously from a single mechanism that arrests and reverses expansion over time.

We identify this mechanism as the \textit{chronophage}\footnote{The term is borrowed
from the ``Chronophage'' or ``time-eater'' atop the Corpus Clock at Corpus Christi
College, Cambridge, a public clock designed to evoke the devouring of time.}: an exponentially accumulating administrative and coordination burden that offsets expansion and drives decline. As systems grow, this burden progressively offsets the gains of expansion, ultimately leading to stagnation and decline. In this sense, the chronophage provides a quantitative realization of the rising costs of complexity, embedding earlier qualitative theories within a predictive dynamical model.

We test this framework on a diverse set of historical empires spanning maritime, land-based, and wartime cases across widely varying time scales. Despite differences in geography, political organization, and historical context, all cases follow a common asymmetric rise–peak–decline trajectory governed by a small number of interpretable parameters. This empirical regularity suggests that the life cycles of large-scale social systems may be governed by universal dynamical constraints. More broadly, the proposed framework identifies the dynamics of empires within a wider class of Gompertz-type processes observed in biological and physical systems of varying size and complexity (Fig. \ref{fig:scales}), including mortality dynamics, tumour growth, microbial populations, and epidemic spread.

\begin{figure}[t] \centering \includegraphics[width=\linewidth]{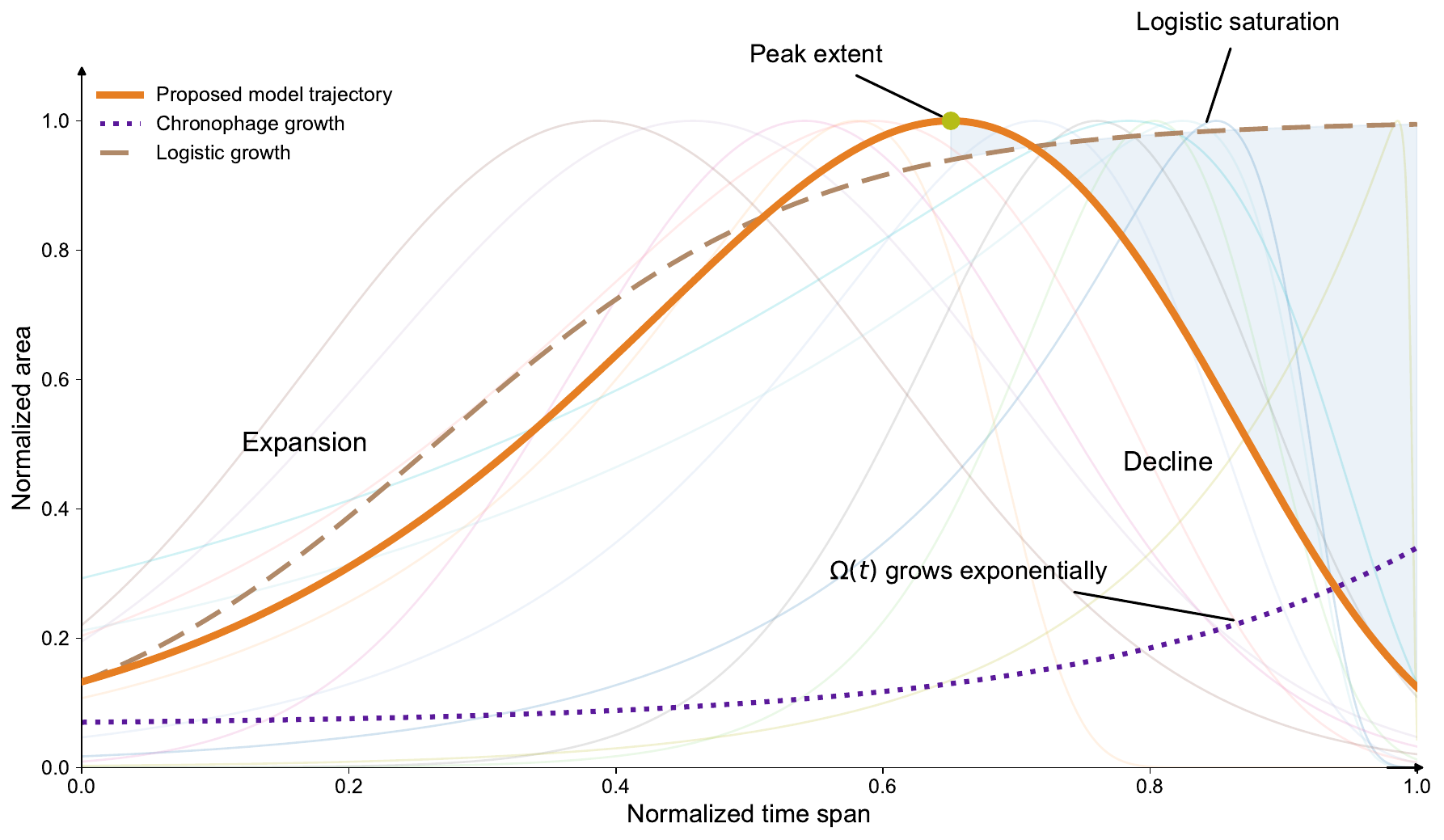} \caption{ \textbf{A common rise--peak--decline trajectory for empires.} Faint curves show normalized empire trajectories with a shared rise–peak–decline pattern. The orange curve is the proposed model; the dashed brown curve is logistic growth without decline. The blue region shows post-peak divergence caused by the chronophage, $\Omega(t)$: the growing administrative burden that offsets expansion and drives contraction. The green marker marks peak territory.}
\label{fig:intro}
\end{figure}

The remainder of the paper proceeds as follows. We first introduce the dynamical framework, specify the proposed rise--peak--decline model, and describe the preprocessing and fitting procedure used for the historical territorial series. We then apply the model to maritime and land-based empires, showing that their territorial histories follow the predicted trajectory. Next, we examine wartime empires, whose expansion and contraction are captured by the same framework despite unfolding on much shorter time scales. Finally, we interpret the fitted parameters in historical terms and discuss the broader implications of the chronophage mechanism for understanding the rise and fall of large territorial systems. Two complementary theoretical derivations are provided in the appendices: \ref{app:maxent} derives the rise--peak--decline envelope from a maximum-entropy construction, while \ref{app:control} shows how the same form emerges from an equivalent optimal-control formulation and its reduced dynamics.

\section{Methods and model}

In this work, we adopt a dynamical-systems approach to model the life cycle of fourteen historical empires: four maritime empires, the British Empire \citep{darwin2009empire}, the French Empire \citep{thomas2014fight}, the Spanish Empire \citep{elliott2006atlantic}, and the Portuguese Empire \citep{disney2009portugal}; six land empires, the Qing Empire and Republic of China \citep{perdue2005china}, the Mongol Empire \citep{morgan2007mongols}, the Islamic Caliphate \citep{kennedy2016prophet}, the Seljuk Empire \citep{peacock2015seljuk}, the Mughal Empire \citep{richards1993Mughal}, and the Delhi Sultanate \citep{jackson1999Delhi}. These ten long-run, Eurasian historical cases were modelled in their expansion phase using a logistic framework by Marchetti and Ausubel \citep{marchetti2012quantitative}; here, however, we broaden the comparison to include four wartime empires: Nazi Germany \citep{Germany1}, Imperial Japan \citep{Japan1}, Napoleonic France \citep{Napoleon1}, and the Italian Empire \citep{Italy1}.

The purpose of the model is to provide a low-dimensional description of the full territorial life cycle of these systems, including both expansion and decline. Although historical empires differ substantially in institutional structure, geography, duration, and external pressures, their territorial trajectories often display a common asymmetric rise--peak--decline pattern \citep{taagepera1978size,taagepera1979size,taagepera1997expansion,arbesman}. The framework developed here treats these trajectories as instances of a generalized Gompertz-like process, in which growth slows exponentially over time and is ultimately offset by increasing constraints. In the context of empires, these constraints are interpreted as the accumulating administrative, fiscal, logistical, and coordination burdens associated with maintaining territorial scale.

\subsection{Proposed model}

The logistic growth equation constitutes a canonical model of bounded, density-dependent expansion and has been extensively employed across the social sciences and quantitative history to describe growth under finite constraints, see Fig. \ref{fig:intro}. In cliodynamic and structural-demographic frameworks \citep{turchin2018historical,sojecka2024global}, logistic-type dynamics often provide the baseline representation of population and aggregate expansion. \cite{marchetti2012quantitative} showed that the territorial
expansion of many historical empires is well approximated by a logistic trajectory,
\[
\dot x = r x(1-x),
\]
where $x(t)$ denotes normalized extent of the area as a function of time $t$, and $r$ is an intrinsic expansion rate.

While
this parsimonious form captures the rise and saturation (peak) of imperial growth, it does not capture the subsequent territorial decline or collapse. This limitation is
structural: a one-dimensional \emph{autonomous} ordinary differential equation (ODE) admits only fixed points as
attractors; trajectories therefore equilibrate monotonically and cannot
generate a single-peaked rise--fall cycle without additional structure. Hence, producing an
asymmetric peak followed by contraction  would require the simplest possible
extension beyond a 1D autonomous model: either (i) a non-autonomous (explicitly
time-dependent) modification or (ii) an equivalent two-dimensional autonomous
system with an additional state variable. We adopt a particularly minimal extension
of the logistic framework by introducing a single time-dependent term governed by
a new parameter, $\alpha$, yielding

\begin{equation}
    \dot x = r x \left(1-e^{\alpha t}\right).
    \label{eq:model}
\end{equation}
~\\
This preserves the logistic-like early expansion but endogenizes a subsequent reversal as the
effective growth factor turns negative, allowing the same closed-form law to capture
the full historical rise--peak--decline trajectory. Equivalently, Eq.~\eqref{eq:model} describes a process in which the instantaneous expansion potential is progressively eroded by a constraint term that grows exponentially through time. We refer to this exponentially accumulating constraint as the \textit{chronophage}: the time-dependent burden of administration, coordination, maintenance, and mobilization that increasingly offsets the gains of expansion.

The model we propose in Eq.~\eqref{eq:model}  admits a closed-form solution that takes the form
\begin{equation}
x(t) = x_{\max}\exp\!\left\{s + s\alpha (t - t_{\max})
- s e^{\alpha (t - t_{\max})} \right\},
\label{eq:solution}
\end{equation}
where $x_{\max}$ denotes the peak territorial extent, $t_{\max}$ the time at which
this peak occurs, $\alpha$ the congestion-growth parameter, and
$s = r/\alpha$ a dimensionless shape parameter controlling the relative steepness
of ascent and decline.

The normalization of territorial series by their historical maxima prior to estimation implies that $x_{\max}$ should lie close to unity across cases, as confirmed by the fitted values reported in Table \ref{tab:params}. The parameters $(r,\alpha)$ determine the intrinsic
expansion rate and the exponential accumulation of mismatch, respectively, while
$s$ summarizes their ratio and thus the overall asymmetry of the trajectory.

Eq.~\eqref{eq:solution} is therefore a generalized Gompertz-like rise--peak--decline envelope. It differs from the standard logistic trajectory by allowing the effective growth rate to become negative after the accumulation of constraints exceeds the expansionary drive. In this sense, the model converts the qualitative idea of diminishing returns to increasing complexity into a simple quantitative mechanism: expansion initially dominates, but the chronophage grows over time, eventually suppressing growth and generating territorial contraction.

\subsection{Data preprocessing}

To facilitate the numerical fit procedure, we regularize each time series within its observed span using a shape-preserving cubic Hermite interpolator (PCHIP). This generates 20 evenly spaced points that pass exactly through the original observations, preserve local monotonicity, avoid spline overshoot, and never extrapolate beyond the data range. The interpolated points thus provide a smooth but faithful proxy trajectory on which the model can be estimated. For transparency, all figures show both the raw data (red) and the interpolated points used for fitting.

The recorded areas \citep{taagepera1997expansion} show major discontinuities at specific times, e.g., the Portuguese Empire losing 90\% of its area between 1820 and 1822 and the Mongolian Empire dropping to 11 million~km$^2$ in 1310 from 24 million~km$^2$ the year before (1309). The Delhi Sultanate series likewise contains an isolated point on the ascending branch that produces an artificial plateau immediately before the maximum. Such implausibly abrupt patterns likely reflect discrete reporting conventions rather than genuine one-step territorial shifts, and their removal yields smoother representations of both imperial expansion and retreat. For France, we add an observation for 1980 equal to metropolitan France plus its overseas departments and territories. This ensures consistency with Taagepera’s convention and makes the late-twentieth-century value directly comparable to the 1962 entry. For the Qing–Republic of China series, we follow Taagepera’s data up to 1920. As a continuation, we add two points: 1949 and 1979, using the area administered by the Republic of China on Taiwan after its relocation. For the wartime cases shown in Fig.~\ref{fig:war}, the territorial series are compiled from the corresponding historical sources \citep{GermanyArea1,JapanArea1,NapoleonArea1,ItalyArea1}.
 
Across all historical cases, the same procedure is followed: raw observations are shown alongside their PCHIP-regularized counterparts, and the proposed model is fitted within the observed span. The resulting curves should be interpreted as stylized trajectories that capture the asymmetric rise–and–fall predicted by the model, while remaining transparent about the sparsity and conventions of the underlying historical data.

For each polity, the territorial area series is normalized by its empirical maximum before fitting. The model in Eq.~\eqref{eq:solution} is then fitted to the PCHIP-regularized trajectory within the observed temporal span. The primary
free parameters controlling the trajectory are $t_{\max}$, $s$, and $\alpha$, with the intrinsic expansion rate recovered as $r=s\alpha$. We also estimate $x_{\max}$ as a scale parameter rather than fixing it to one, using it as a sanity check on the normalization; because the territorial series are normalized
by their empirical maxima, fitted values of $x_{\max}$ are expected to remain close to unity. Thus, the effective shape of the trajectory is governed by three parameters, while $x_{\max}$ captures residual scale mismatch.

Goodness of fit is summarized using the root mean squared error (RMSE) between the fitted curve and the interpolated normalized series. The same fitting procedure is applied to all maritime, land-based, and wartime cases, allowing differences across empires to be expressed through parameter values rather than through changes in functional form.

\begin{figure}[t]
\captionsetup[subfigure]{labelformat=empty}
  \centering
  \subfloat[][(a) British Empire]{\includegraphics[width=.48\textwidth]{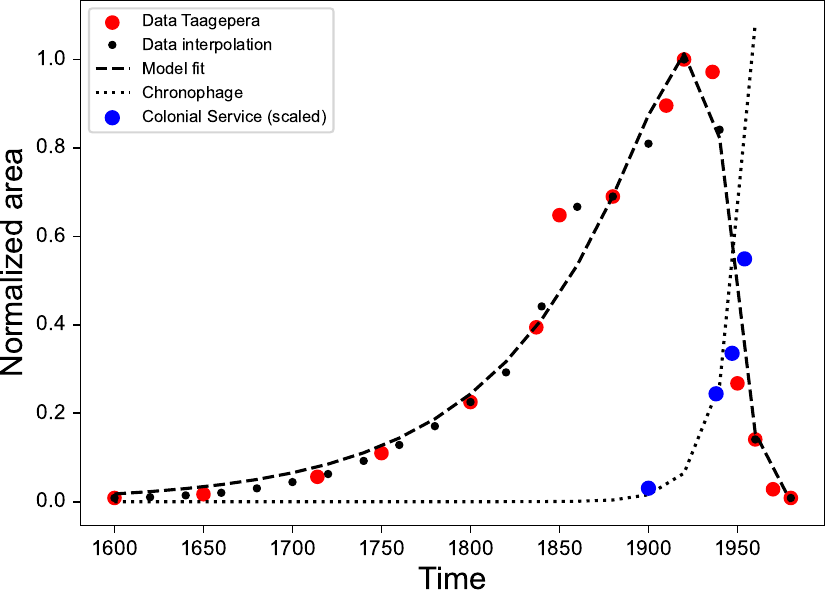}}\quad
  \subfloat[][(b) French Empire]{\includegraphics[width=.48\textwidth]{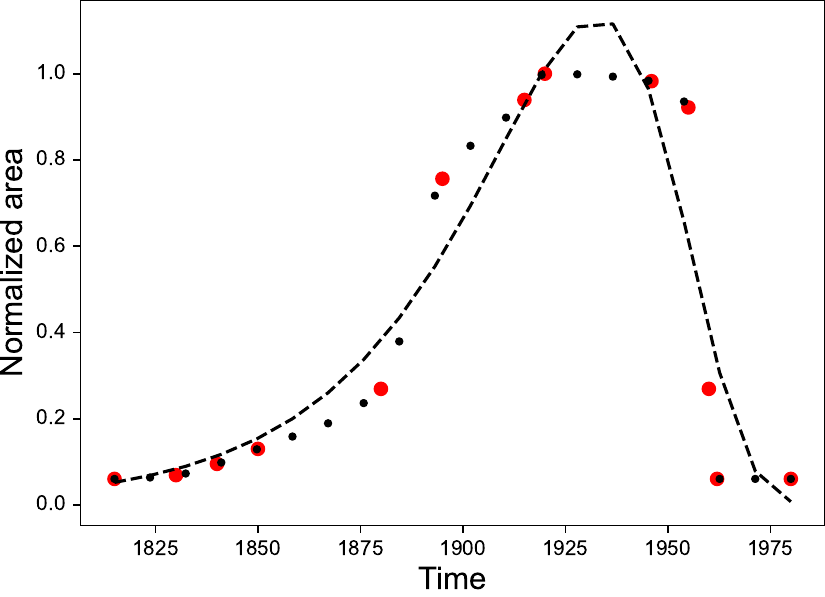}}\\
  \subfloat[][(c) Spanish Empire]{\includegraphics[width=.48\textwidth]{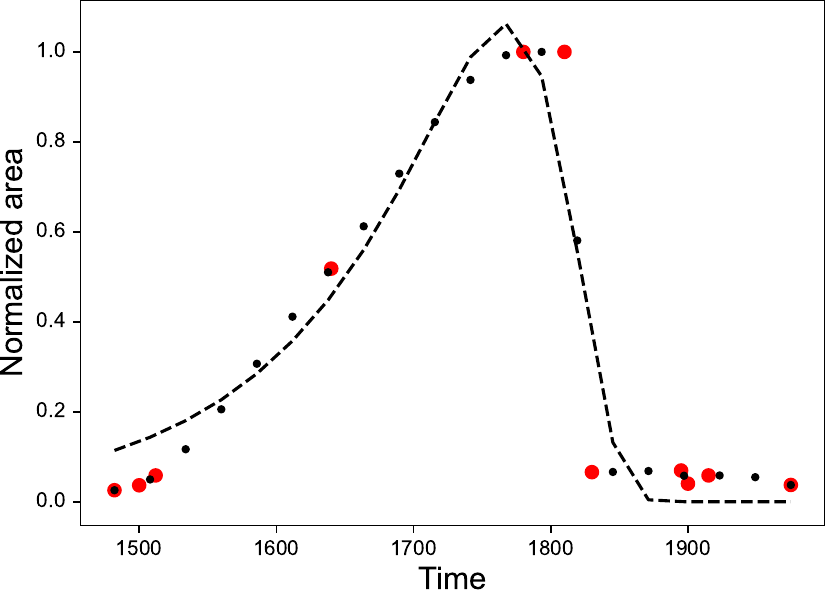}}\quad
  \subfloat[][(d) Portuguese Empire]{\includegraphics[width=.48\textwidth]{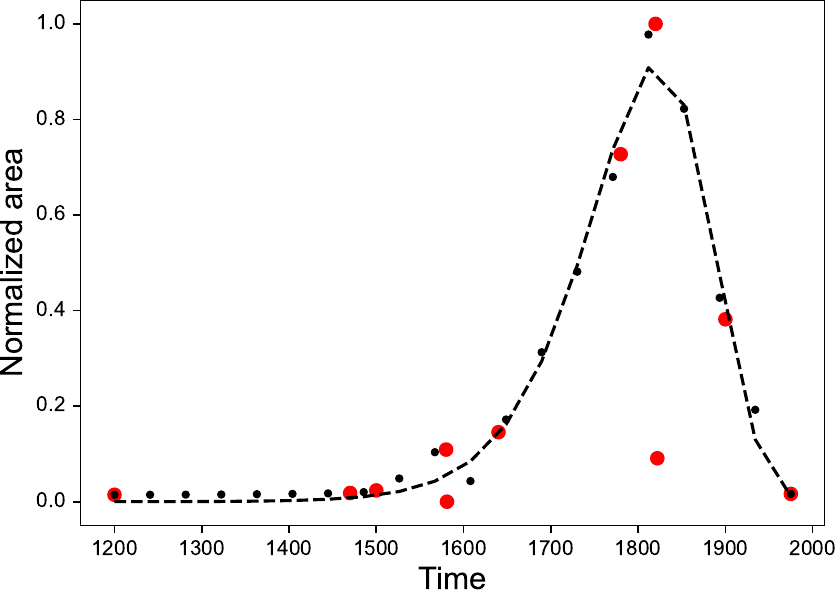}}
  \caption{The fit of the proposed model to the territorial expansion of maritime empires. Red dots denote original data \citep{taagepera1978size,taagepera1979size,taagepera1997expansion}; black curves show fits of the proposed model estimated on shape-preserving interpolated series. In (a) British Empire, the blue dots represent the chronophage proxy data, scaled appropriately for visual clarity.}
  \label{fig:sea}
\end{figure}

\section{Results}

The territorial trajectories of the major maritime empires (British, French, Spanish and Portuguese) are well described by Eq.~\eqref{eq:solution} under a three-parameter shape fit, with $x_{\max}$ retained as a scale check. As shown in Fig.~\ref{fig:sea}, the fitted curve reproduces the characteristic empirical pattern: a sustained, approximately logistic rise, a single interior maximum at $t_{\max}$, and a prolonged, asymmetric decline. Within Eq.~\eqref{eq:solution}, early expansion is governed primarily by the linear growth component, whereas the nested exponential term progressively suppresses net growth, generating an endogenous turning point and a right-skewed retreat. Despite substantial heterogeneity in temporal features and scale, cross-case differences are captured parsimoniously through variation in $(t_{\max}, s, \alpha)$, while the functional form remains invariant. This suggests that imperial expansion conforms closely to a common rise–peak–decline envelope.

\begin{figure}[]
\captionsetup[subfigure]{labelformat=empty}
  \centering
  \subfloat[][(a) Qing Dynasty-ROC]{\includegraphics[width=.48\textwidth]{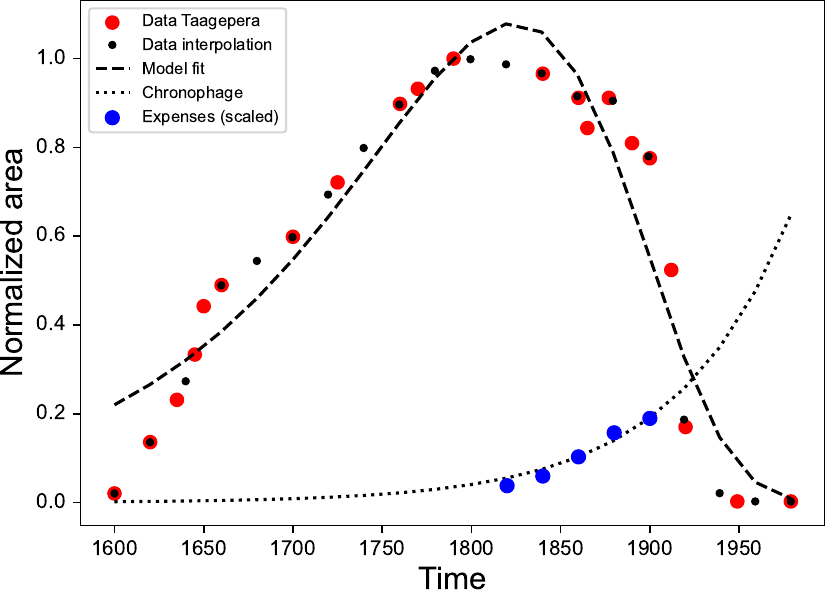}}\quad
  \subfloat[][(b) Mongol-Y\"uan Empire]{\includegraphics[width=.48\textwidth]{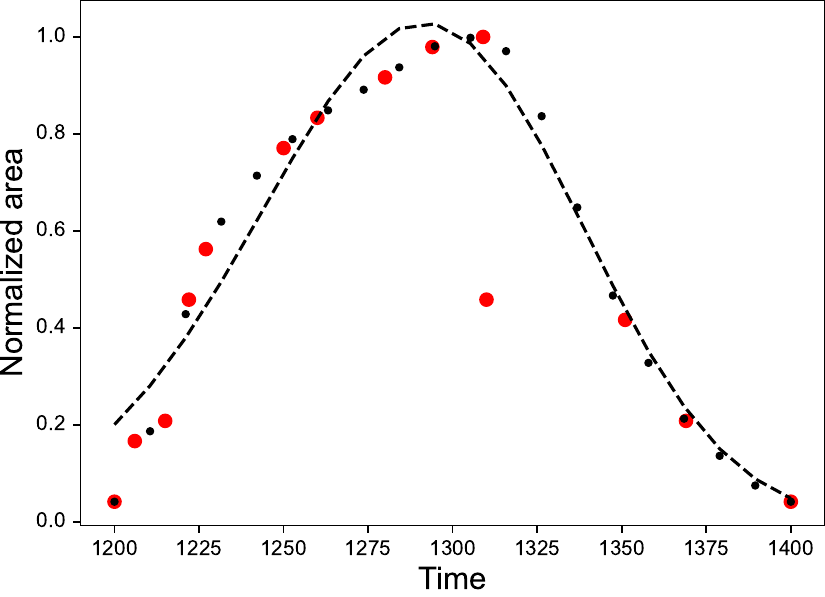}}\\
  \subfloat[][(c) Islamic Caliphate]{\includegraphics[width=.48\textwidth]{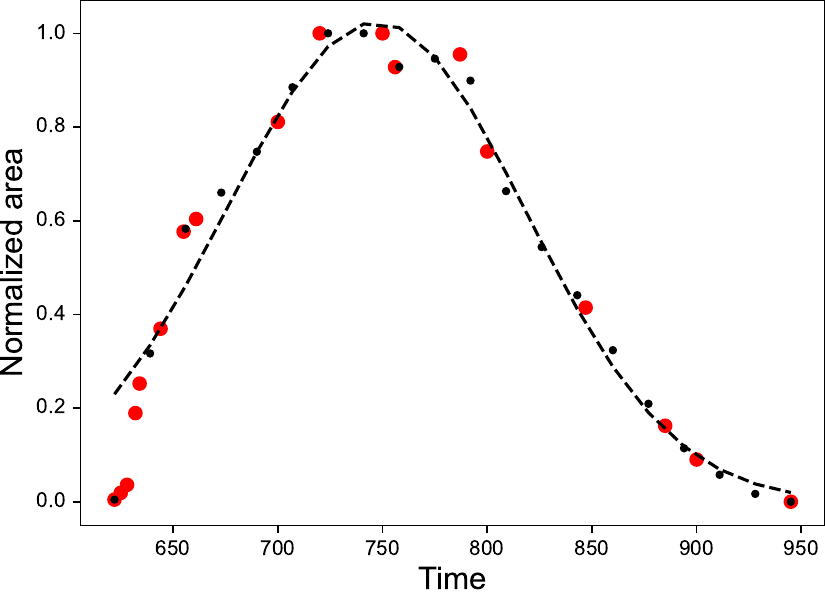}}\quad
  \subfloat[][(d) Seljuk Empire]{\includegraphics[width=.48\textwidth]{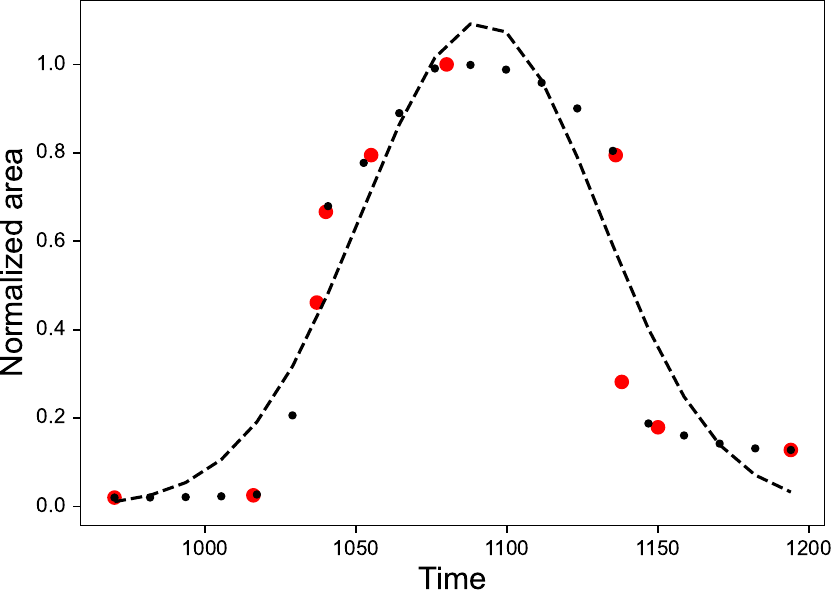}}\\
  \subfloat[][(e) Mughal Empire]{\includegraphics[width=.48\textwidth]{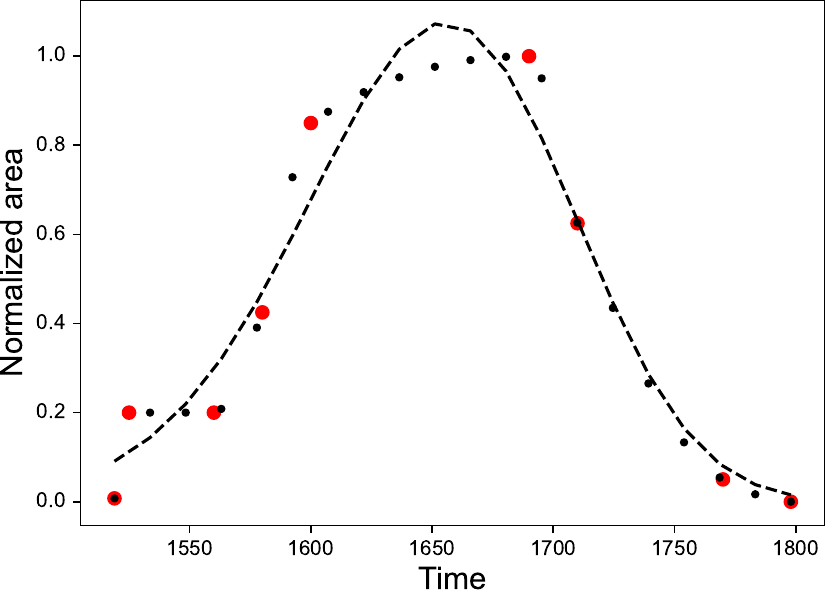}}\quad
  \subfloat[][(f) Delhi Sultanate]{\includegraphics[width=.48\textwidth]{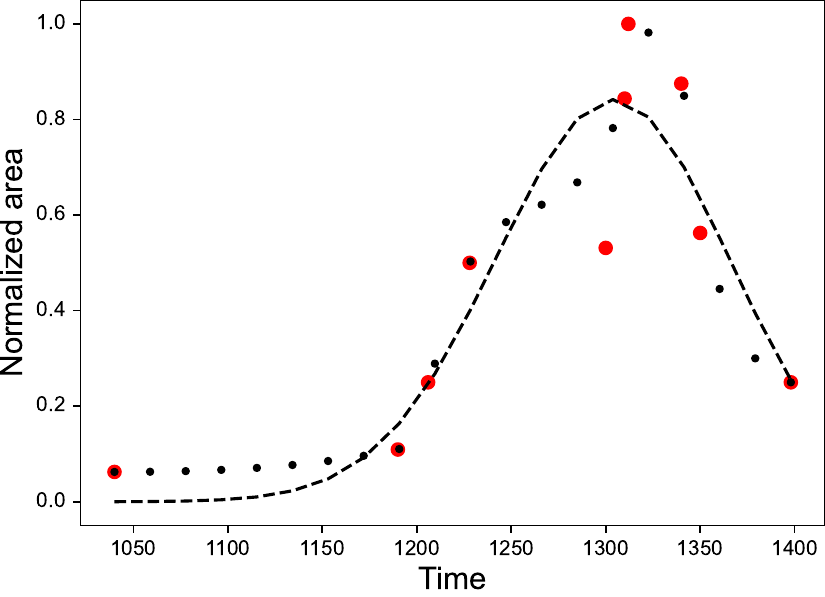}}
  \caption{The fit of the proposed model to the territorial expansion of land-based empires. In (a) Qing Dynasty-ROC, the blue dots represent the chronophage proxy data, scaled appropriately for visual clarity.}
  \label{fig:land}
\end{figure}

A similar result holds for major land-based empires (Mongol, Qing--Republic of China (ROC), Caliphate, Seljuk, Mughal and Delhi). As shown in Fig.~\ref{fig:land}, Eq.~\eqref{eq:solution} provides an equally close fit to their territorial histories, capturing both rapid pre-peak expansion and extended post-peak contraction. Differences across cases are reflected primarily through parameter values (see Table \ref{tab:params}), most notably the timing of the maximum and the relative strength of the congestion term, rather than through changes in functional structure. In all cases, the same mechanism encoded in Eq.~\eqref{eq:solution} produces a single peak followed by asymmetric decline, indicating that maritime and land empires share a common quantitative growth profile despite distinct geopolitical contexts.

We have seen so far that decline, rather than stability, is a common phenomenon that follows growth. To elucidate the mechanism underlying this process of decline, the model in Eq.~\eqref{eq:model} can be rewritten equivalently as an autonomous system of two coupled ordinary differential equations:
\begin{equation}
    \dot{x}(t) = x(t)\bigl[s\alpha - \Omega(t)\bigr], \qquad
\dot{\Omega}(t) = \alpha\,\Omega(t).
\label{eq:2d}
\end{equation}
In this form the model resembles
a predator--prey system in which territorial extent $x(t)$ is suppressed by a
``predator'' $\Omega(t)$ that grows autonomously and, unlike in the case of classical
Lotka--Volterra dynamics, does not decay. This one-way coupling produces a single
rise--peak--decline trajectory rather than oscillations, consistent with the historical patterns described above. The variable $\Omega(t) \sim e^{\alpha t} $ represents the chronophage, an exponentially growing stock of
administrative congestion or coordination burden and substitutes the single-time dependent term from the prior formulation in Eq. \eqref{eq:model}.

The chronophage mechanism encoded in the parameter $\alpha$ can be reflected in concrete historical trends, relating to the multiplicative growth of coordination, monitoring and maintenance burdens as territorial systems grow. Two distinct historical proxies, administrative personnel in the British Empire (see Fig.~\ref{fig:sea}a) and fiscal expansion in the Qing dynasty (see Fig.~\ref{fig:land}a) illustrate how such congestion can accumulate at rates commensurate with the estimated parameter values in Eq.~\eqref{eq:solution}. 

In the late British Empire, the size of the Colonial Service expanded rapidly, rising from roughly $1{,}000$ overseas officers around 1900 to approximately
$8{,}000$ by 1938, about $11{,}000$ by 1947, and nearly $18{,}000$ posts by 1954 \citep{kirk1980thin}. Approximated as exponential growth, this corresponds to an
annual increase on the order of $5\text{--}7\%$. This magnitude is comparable to the fitted congestion parameter for Britain ($\alpha \approx 7.05 \times
10^{-2}\,\mathrm{yr}^{-1}$). Within the chronophage framework, the rapid expansion of overseas administrative personnel reflects escalating demands for
governance, information processing and supervision across a geographically dispersed empire. As these demands compound, they absorb an increasing share of resources and progressively offset the gains from further territorial
acquisition, contributing to the asymmetric post-peak decline generated by Eq.~\eqref{eq:solution}.

A parallel dynamic appears in the Qing dynasty, where the chronophage mechanism is reflected in rising fiscal strain. Reconstructed Qing fiscal series show that during the nineteenth century state expenditures grew
persistently relative to revenues, pushing the fiscal balance from surplus toward sustained deficit \citep{orlandi2023structural}. This deterioration is
primarily attributed to the mounting costs of large-scale internal rebellions and associated military mobilization, which imposed prolonged upward pressure on state outlays. Expressed in growth terms, total expenditures increased at
approximately 2\% per year over extended late-imperial intervals, a rate consistent with the fitted chronophage parameter reported in Table \ref{tab:params}. Within the framework of Eq.~\eqref{eq:solution}, this correspondence supports the interpretation of $\alpha$ as capturing the accumulation of systemic maintenance burdens required to preserve territorial integrity and internal order.

\begin{figure}[t]
\captionsetup[subfigure]{labelformat=empty}
  \centering
  \subfloat[][(a) Nazi Germany]{\includegraphics[width=.48\textwidth]{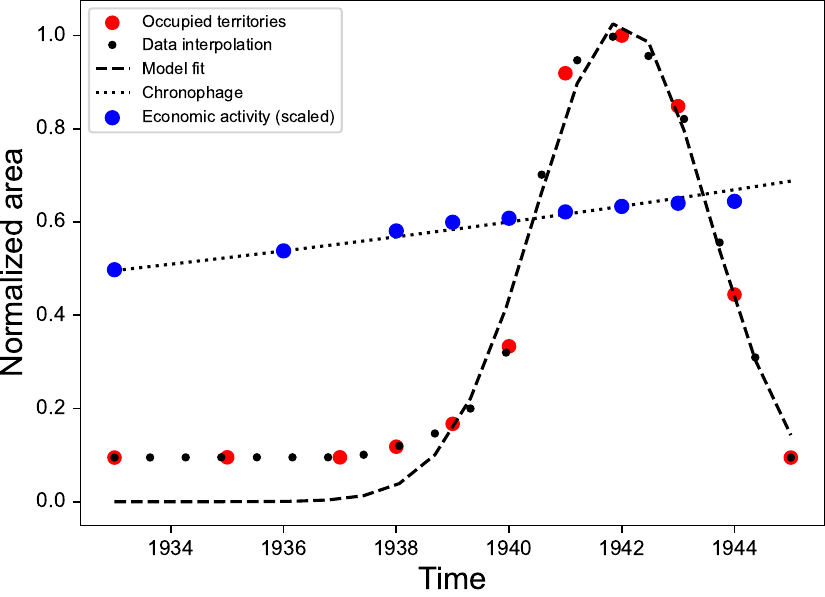}}\quad
  \subfloat[][(b) Imperial Japan]{\includegraphics[width=.48\textwidth]{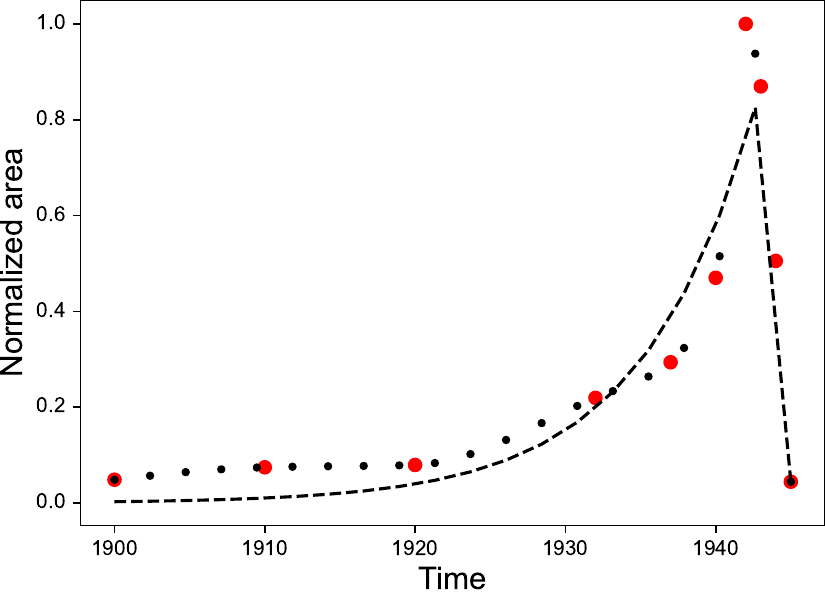}}\\
  \subfloat[][(c) Napoleonic France]{\includegraphics[width=.48\textwidth]{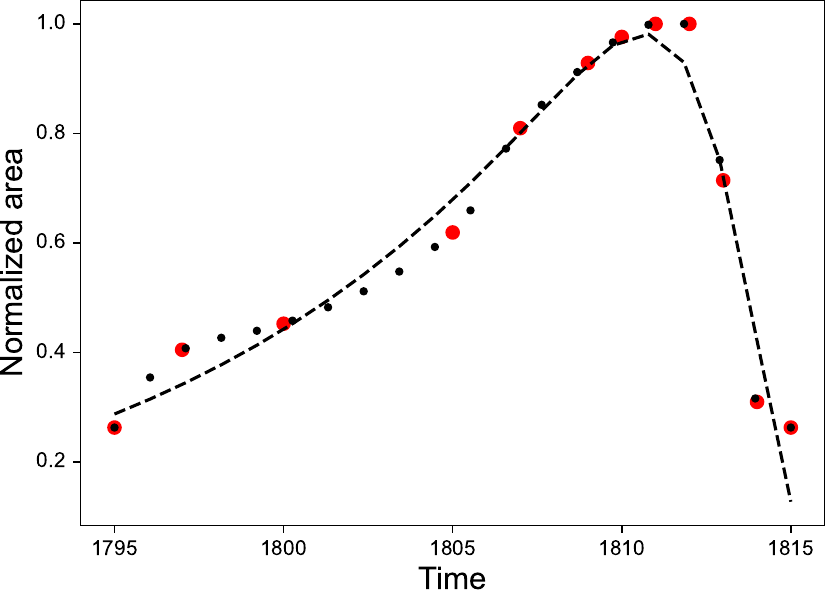}}\quad
  \subfloat[][(d) Italian Empire]{\includegraphics[width=.48\textwidth]{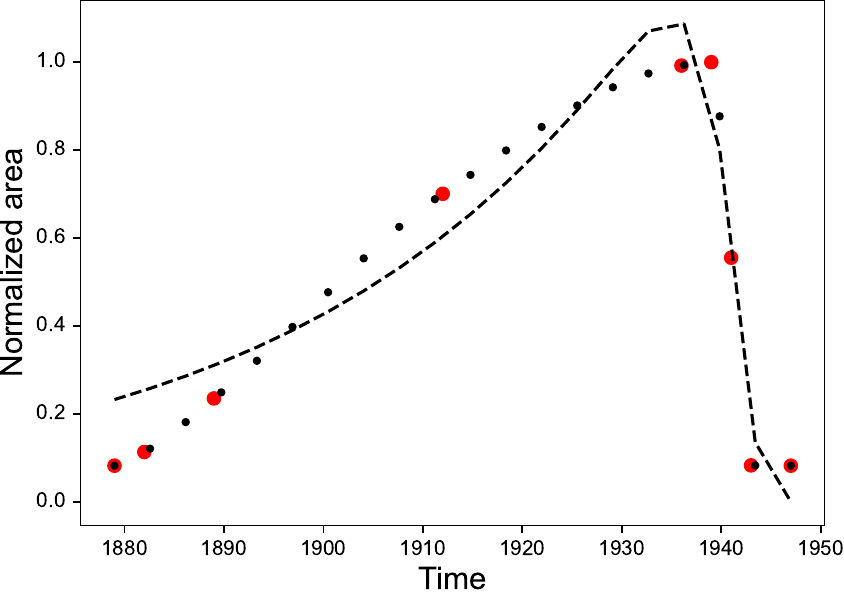}}
  \caption{Fit of the proposed model to wartime territorial expansions of empires. Red dots denote data on occupied territories for each wartime case. In (a) Nazi Germany, the blue dots represent the chronophage proxy data, scaled appropriately for visual clarity.}
  \label{fig:war}
\end{figure}

The functional form of Eq.~\eqref{eq:solution} also describes a class of short-lived, highly expansionary polities whose territorial growth was closely tied to major military conflicts. Fig.~\ref{fig:war} shows four such cases with data available from associated sources: Nazi Germany \citep{GermanyArea1}, Imperial Japan \citep{JapanArea1}, Napoleonic France \citep{NapoleonArea1} and the Italian Empire \citep{ItalyArea1}. In contrast to the maritime and land empires discussed above, whose expansion unfolded over centuries, these polities expanded rapidly during periods of intense warfare and collapsed within a few years. Despite this difference in time scales, their territorial trajectories exhibit a structural pattern akin to the cases discussed above: a steep ascent, a peak located near the end of the expansion phase, and a rapid but asymmetric contraction. Eq.~\eqref{eq:solution} captures these dynamics using the same minimal parameterization as in the historical cases. The main difference lies in the parameter values (see Table \ref{tab:params}): wartime empires often exhibit substantially larger expansion rates $r$, reflecting accelerated conquest dynamics, while the chronophage parameter $\alpha$ remains positive and produces the same endogenous turning point. The persistence of the same rise–peak–decline envelope across these very different historical contexts suggests that the mechanism encoded in the model is not limited to long administrative processes but can also govern rapid geopolitical expansions driven by military mobilization.

For Nazi Germany, the chronophage can be interpreted through the dynamics of the wartime economy. As a proxy for the scale of economic activity, Fig.~\ref{fig:war}a plots the square root of German GDP during the late 1930s and early 1940s \citep{GermanyGDP1}. The use of the square root follows a simple mass-action intuition: if aggregate output arises from interactions within the productive system, total GDP scales approximately with the square of the economy’s effective degree of interaction. Taking the square root of GDP therefore provides an estimate of this underlying interaction. When expressed in this way, the resulting series grows at roughly three percent per year over the relevant interval, closely matching the chronophage growth rate reported for Nazi Germany in Table \ref{tab:params}. Within this interpretation, the chronophage represents the increasing share of economic capacity absorbed by mobilisation, occupation administration and wartime logistics. Unlike the long-run imperial cases discussed above, where the chronophage reflects gradually accumulating coordination burdens, here it acts as an economic bottleneck to political and military ambitions, constraining further territorial expansion and contributing to the abrupt decline that follows the wartime peak.
\begin{table}[t]
\centering
\caption{Fitted parameters for each polity under the proposed model. Span gives the data window used for fitting. $t_{\max}$ and $x_{\max}$ are in standard decimal; all other quantities use three significant digits in scientific notation. RMSE is the Root Mean Squared Error of the fit.}
\label{tab:params}
\small
\setlength{\tabcolsep}{4pt}
\renewcommand{\arraystretch}{1.12}
\begin{tabular}{lccccccc}
\hline
Polity & Span & $t_{\max}$ & $x_{\max}$ & $s$ & $\alpha$ & $r=s\alpha$ & RMSE \\
\hline
British Empire & 1600--1980 & 1923 & 1.02  & $1.87 \times 10^{-1}$ & $7.05 \times 10^{-2}$ & $1.31 \times 10^{-2}$ & $3.69 \times 10^{-2}$ \\
French Empire & 1815--1980 & 1933 & 1.13  & $5.92 \times 10^{-1}$ & $5.24 \times 10^{-2}$ & $3.10 \times 10^{-2}$ & $1.09 \times 10^{-1}$ \\
Spanish Empire & 1482--1975 & 1768 & 1.06  & $2.91 \times 10^{-1}$ & $3.04 \times 10^{-2}$ & $8.83 \times 10^{-3}$ & $5.60 \times 10^{-2}$ \\
Portuguese Empire & 1200--1975 & 1823 & 0.92 & $1.63 \times 10^{0}$  & $1.10 \times 10^{-2}$ & $1.80 \times 10^{-2}$ & $3.37 \times 10^{-2}$ \\
Qing Dynasty–ROC & 1600--1979 & 1824 & 1.08  & $6.36 \times 10^{-1}$ & $1.55 \times 10^{-2}$ & $9.84 \times 10^{-3}$ & $1.02 \times 10^{-1}$ \\
Mongol-Y\"uan Empire & 1200--1400 & 1292 & 1.03  & $2.44 \times 10^{+1}$  & $4.25 \times 10^{-3}$ & $1.04 \times 10^{-1}$ & $6.59 \times 10^{-2}$ \\
Islamic Caliphate  & 622--945   & 747 & 1.02  & $6.92 \times 10^{+2}$  & $5.31 \times 10^{-4}$ & $3.67 \times 10^{-1}$ & $6.55 \times 10^{-2}$ \\
Seljuk Empire & 970-–1194  & 1092 & 1.10  & $6.48 \times 10^{+2}$  & $9.98 \times 10^{-4}$ & $6.47 \times 10^{-1}$ & $1.09 \times 10^{-1}$ \\
Mughal Empire & 1519--1798 & 1655 & 1.08 & $1.37 \times 10^{+1}$ & $4.90 \times 10^{-3}$ & $6.69 \times 10^{-2}$ & $7.01 \times 10^{-2}$ \\
Delhi Sultanate & 1040--1398 & 1315 & 0.87 & $1.28 \times 10^{0}$ & $1.58 \times 10^{-2}$ & $2.03 \times 10^{-2}$ & $7.21 \times 10^{-2}$ \\
Nazi Germany & 1933--1945 & 1942 & 1.03 & $5.81 \times 10^{+2}$ & $2.73 \times 10^{-2}$ & $1.59 \times 10^{+1}$ & $6.97 \times 10^{-2}$ \\
Imperial Japan & 1895--1945 & 1944 & 1.02 & $1.68 \times 10^{-2}$ & $8.00 \times 10^{0}$ & $1.34 \times 10^{-1}$ & $6.00 \times 10^{-2}$ \\
Napoleonic France & 1795--1815 & 1811 & 0.98 & $1.23 \times 10^{-1}$ & $7.01 \times 10^{-1}$ & $8.62 \times 10^{-2}$ & $5.32 \times 10^{-2}$ \\
Italian Empire & 1882--1947 & 1935 & 1.10 & $6.76 \times 10^{-2}$  & $4.27 \times 10^{-1}$ & $2.89 \times 10^{-2}$ & $8.21 \times 10^{-2}$ \\
\hline
\end{tabular}
\end{table}

\section{Discussion}

For the case of historical maritime and land empires (red points in Fig.~\ref{fig:sea} and Fig.~\ref{fig:land}), we use data from Taagepera’s compilations \citep{taagepera1997expansion,taagepera1978size}. A common feature of these datasets is the convention of recording the maximum territorial extent immediately prior to a major contraction. While consistent across empires and convenient for encyclopedic reference, this practice tends to accentuate abrupt breaks and understate the continuity of more gradual adjustments in influence, administration, and economic reach. As a result, declines may appear as sudden collapses rather than the drawn-out retreats they often were. The discrepancy between these step-like observations and the smoother dynamics implied by the model is visible, for example, in the Seljuk decline. To mitigate this effect, we smooth a small number of clear discontinuities, most notably in the Portuguese and Mongolian series, so that the fitted trajectories capture longer-run dynamics rather than isolated reporting steps; details are provided in the Methods section. In doing so, this can additionally introduce artificial plateaus or transient humps in some cases, such as the British, Qing, and Napoleonic empires, which should be interpreted as artifacts of the underlying data construction rather than substantive features of the dynamics.

Following the pre-processing, the functional form of Eq.~\eqref{eq:solution} is fitted to the data. All resultant parameters, stated in Table \ref{tab:params}, have direct interpretations in terms of observable
quantities. The parameter $t_{\max}$ identifies the timing of the maximum area and aligns closely with the historically recognized apex of each polity.
The intrinsic rate $r$ governs the pace of territorial expansion during the early
phase of growth, while $\alpha$ measures the rate at which the limiting mechanism
represented by the chronophage accumulates over time. The dimensionless parameter
$s=r/\alpha$ therefore captures the balance between expansionary momentum and the
speed at which systemic constraints emerge. Together, these quantities map the
empirical trajectories of territorial extent onto interpretable measures of timing,
growth and decline.

Across the maritime and long-run land empires shown in Fig.~\ref{fig:sea} and Fig.~\ref{fig:land}, the fitted parameters reveal several distinct regimes. The maritime empires, together with the Qing and Delhi cases, exhibit intrinsic expansion rates of a few percent per unit time despite large differences in era and geography. By contrast, conquest-dominated land empires such as the Mongol, Caliphate, Seljuk, and Mughal cases occupy a markedly faster-growth regime, with substantially larger $r$ than the maritime cases. These differences in $r$ coexist with systematic variation in $\alpha$: maritime empires typically exhibit relatively larger $\alpha$, indicating that the limiting mechanism accumulates more rapidly in overseas systems than in most contiguous land empires, although Qing and Delhi lie closer to the maritime end of this spectrum. Because $s=r/\alpha$, similar qualitative rise--fall shapes can therefore arise from very different decompositions into expansion intensity ($r$) and constraint accumulation ($\alpha$). In this sense, high $r$ is not unique to modern warfare, but also characterizes several historical episodes of rapid territorial consolidation across the land empires considered here.

However, wartime polities in Fig.~\ref{fig:war} do not form a unified parameter
regime. Their fitted $r$ values are generally higher than those of the maritime
colonial cases, but overlap with the above mentioned conquest-dominated land empires. Moreover,
$\alpha$ varies widely across wartime cases: some exhibit very rapid constraint
growth (large $\alpha$), while others do not, precluding the classification of ``wartime” empires as a uniform high-$\alpha$ category. Instead, the consistency is
structural and temporal: these trajectories are compressed into short spans,
with peaks $t_{\max}$ occurring close to the end of the expansion window and
contraction following quickly thereafter. In this setting the chronophage
should be interpreted less as slowly accumulating coordination burdens and more
as rapidly binding constraints associated with mobilisation, occupation and
multi-front war, which can tighten on the same short horizon as territorial
expansion. 
Although there are marked differences between the cases, the same functional form captures both long-run imperial expansion and short wartime territorial surges, with variation across cases expressed primarily through parameter values rather than through changes in the underlying dynamical mechanism.

%\subsection*{Alternate model interpretations}

At a fundamental level, the functional form can be motivated from an
information-theoretic perspective. As shown in \ref{app:maxent}, among unimodal asymmetric
envelopes consistent with a fixed mean and an exponential moment reflecting
the accumulation of multiplicative hazards or coordination burdens, the
Gompertz-type curve arises as the maximum-entropy solution \citep{banavar2010applications,roman2025maximum}. In this
interpretation the trajectory represents the minimum-bias curve compatible
with the assumed structural constraints. The resulting envelope therefore
introduces no additional structure beyond what is required by the constraints
themselves, providing a neutral baseline against which the empirical rise and
fall of large territorial systems can be analysed and interpreted.

This maximum-entropy envelope also possesses several practical advantages for
empirical analysis. Imperial trajectories are typically asymmetric, with a
relatively gradual phase of expansion followed by a more rapid and skewed
decline. The envelope implied by Eq.~\eqref{eq:solution} reproduces this
pattern naturally, generating an approximately exponential rise followed by a
faster contraction without requiring additional ad hoc parameters. At the same
time, the formulation remains analytically simple and parsimonious: only a
small number of parameters determine the scale, timing and asymmetry of the
trajectory, and each maps directly onto interpretable quantities such as the
expansion rate, the timing of the peak and the rate at which systemic burdens
accumulate. Alternative unimodal curves, including Richards, skew--normal,
generalized Gamma or Beta forms, can also reproduce asymmetric trajectories but typically either
require additional shape parameters or lack clear interpretability, making
cross-case comparison more difficult.

The same trajectory emerges naturally from simple dynamical formulations. In the reduced chronophage representation introduced
above, territorial extent evolves under the influence of a burden that grows
exponentially and gradually suppresses expansion, yielding a single
rise--peak--decline cycle. From an economic perspective, the same envelope arises in an optimal-control framework in which the empire balances the gains from territorial expansion against the costs of sustaining support capacity. As detailed in \ref{app:control}, territorial extent and support capacity evolve as coupled state variables, while imperial policy adjusts endogenously under expansion costs. Under this specification, the optimal policy generates a trajectory whose early transient reproduces the dynamics of Eq.~\eqref{eq:model}. Expansion initially dominates but is later constrained as coordination costs accumulate and territorial reach outpaces support capacity.

To clarify the historical context in which the framework developed so far can reasonably apply, we identify a set of premises under which a single rise–peak–decline cycle can be meaningfully represented by a unified dynamical law. Over the fitted span we therefore require:

\begin{itemize}
\item \textbf{P1. A single coherent rule set.} The polity operates under a broadly consistent strategic and institutional framework without major regime resets that would change the governing parameters within the fitted window.
\item \textbf{P2. Aggressive and sustained expansion.} Territorial growth is deliberate and persistent rather than opportunistic or episodic.
\item \textbf{P3. Local dominance during ascent.} The historical evolution is primarily determined by internal capacities rather than sustained subordination to stronger external rivals in the principal theatre.
\item \textbf{P4. Early cheap-growth orientation.} Initial territorial gains are pursued as if they can be incorporated with relatively limited increases in administrative, fiscal, or logistical capacity.
\end{itemize}
These premises together define a regime in which expansion initially proceeds rapidly but gradually generates internal coordination burdens that eventually reverse the trajectory.

The historical maritime and land empires examined here satisfy these conditions to a sufficient degree such that their rise and decline can be described by an autonomous dynamical system driven primarily by internal mechanisms. The British, French, Spanish, and Portuguese empires each pursued sustained overseas expansion under relatively stable administrative settings and enjoyed extended periods of naval or geopolitical advantage during their ascent. Similarly, the Mongol, Caliphate, Seljuk, Qing--ROC, and Mughal trajectories reflect coherent imperial projects characterized by purposeful territorial enlargement and early phases in which expansion occurred with comparatively low resistance. Under these conditions, the rise and subsequent decline can be interpreted as emerging endogenously from the interaction between expansionary drive and the gradual accumulation of coordination burdens, producing a single asymmetric expansion--contraction cycle. The Delhi Sultanate is more borderline: its trajectory can be treated within the same framework only if dynastic succession is understood not as a sequence of regime resets, but as preserving a sufficiently continuous outward imperial orientation over the fitted span, with later fragmentation interpreted as the decline of that same project; Timur's invasion then acts less as a separate cycle than as an exogenous accelerator of an already weakening trajectory.

\begin{table}[]
\centering
\caption{Empires that do not satisfy one or more premises required for the single-cycle dynamics assumed by the proposed model.}
\label{tab:counterexamples}

\small
\renewcommand{\arraystretch}{0.97}
\setlength{\tabcolsep}{4pt}
\setlength{\aboverulesep}{0pt}
\setlength{\belowrulesep}{0pt}
\setlength{\extrarowheight}{0pt}

\begin{tabularx}{\linewidth}{
>{\raggedright\arraybackslash}p{2.8cm}
>{\raggedright\arraybackslash}p{2.2cm}
>{\raggedright\arraybackslash}X
}
\toprule
\textbf{Polity} & \textbf{Failed premises} & \textbf{Reason} \\
\midrule

Dutch Empire & P1, P2, P3, P4 &
A dispersed commercial empire organised around chartered companies and trade networks rather than a unified territorial project. Expansion was episodic and profit-driven, local dominance was contested by rival maritime powers, and growth depended on commercial concessions rather than territorial integration. \\
\addlinespace[2pt]

Danish Empire & P1, P2, P3 &
Colonial possessions were geographically scattered and strategically secondary to the metropolitan state. Expansion was limited and opportunistic rather than sustained, and Denmark rarely exercised clear regional dominance in the imperial theatre. \\
\addlinespace[2pt]

Roman Empire & P4 &
Expansion relied on a highly integrated fiscal--military and administrative system that scaled with territorial reach. Frontier consolidation through taxation, infrastructure, and permanent forces implies expansion was not treated as low-friction growth, and the long plateau preceding decline reflects strong internal feedback mechanisms. \\
\addlinespace[2pt]

Holy Roman Empire & P1, P2, P3 &
A polycentric and elective composite polity rather than a unified strategic project, lacking sustained aggressive expansion and centralised territorial dominance. Its structure prevents treatment as a single coherent expansionary trajectory. \\
\addlinespace[2pt]

Byzantine Empire & P1, P2, P3 &
Repeated institutional and military reorganisations reset governing parameters, while the empire alternated between contraction and partial recovery under persistent pressure from peer powers. The resulting trajectory contains multiple cycles rather than a single rise--peak--decline pattern. \\
\addlinespace[2pt]

Ottoman Empire & P1, P3 &
The empire underwent major administrative and institutional transformations across its long history, and expansion occurred in sustained competition with powerful regional rivals. These factors prevent treating the trajectory as a single internally driven cycle. \\
\addlinespace[2pt]

Safavid Persia & P1, P3, P4 &
Repeated military and administrative reorganisations, persistent peer pressure, and high campaign costs prevent treatment as a single cheap--growth expansion. \\
\addlinespace[2pt]

Ming Empire & P2, P3 &
Imperial strategy emphasised consolidation and internal governance rather than sustained outward territorial expansion. Regional rivals and frontier dynamics constrained dominance, producing a trajectory marked by plateaus and not just expansion or decline. \\

\bottomrule
\end{tabularx}
\end{table}

The wartime cases operate under the same structural premises \textbf{P1-P4}, albeit in a markedly different geopolitical environment. The expansionary ambitions of regimes such as Nazi Germany, Imperial Japan, Napoleonic France, and the Italian Empire rapidly provoked coordinated responses from rival powers. As a result, the dynamics cannot be interpreted as purely autonomous: external resistance becomes progressively stronger as expansion proceeds. Nevertheless, these reactions feed back into the same internal mechanism represented in Eq.~\eqref{eq:model}. Military commitments, administrative burdens, and economic mobilization increase the demands of maintaining territorial control, amplifying the internal constraints that limit expansion. The trajectory therefore reflects a reinforcing spiral in which expansion catalyzes international opposition, which in turn raises the internal costs of sustaining the enlarged system and accelerates the transition from rapid ascent to abrupt decline.

However, not all historical empires necessarily satisfy all the premises \textbf{P1-P4} concomitantly. Table \ref{tab:counterexamples} illustrates how these premises help delimit the historical situations in which the proposed model provides a meaningful description. The empires listed share certain superficial similarities with the fitted cases, including large territorial scale, long duration, or participation in the same maritime or Eurasian geopolitical systems, yet their trajectories do not correspond to the single internally driven expansion--contraction cycle assumed by the model. In several cases, such as the Dutch and Danish empires \citep{Gaastra2003}, the difficulty arises because expansion was organised primarily through decentralized commercial networks rather than a unified territorial project.

In others, including the Byzantine \citep{Shepard2008} and Holy Roman \citep{StollbergRilinger2018} empires, repeated institutional transformations and persistent competition with comparable powers prevented the emergence of a single coherent expansionary trajectory. Large land empires such as Rome \citep{roman2019growth} or the Ottomans also illustrate additional limitations. Their long histories involved substantial administrative adaptation, multiple strategic phases, and extended plateaus that are historically meaningful, and are better interpreted as outcomes of complex feedback processes than of a single expansionary pulse. The Ming case highlights a different issue, where territorial dynamics reflect long phases of consolidation within a broader sequence of dynastic transitions \citep{roman2021historical}. The Safavid Empire likewise does not satisfy the premises of the model, as its territorial trajectory was shaped by sustained rivalry with the Ottoman Empire and ultimately ended through external conquest during the Afghan invasion of the early eighteenth century \citep{Newman2006}. These examples thus go beyond highlighting discrepancies and instead contribute to clarifying the scope of the model. In summary, the framework developed in this work applies to cases where territorial growth and contraction form a relatively coherent imperial project governed primarily by internal dynamics rather than repeated institutional resets or externally imposed discontinuities.

\section{Conclusion}

In this work, we propose a minimal extension of logistic growth that captures not only
the expansion of empires but also their decline. Across cases of various historical maritime, land-based and
wartime empires, the same closed-form law reproduces the characteristic observed
pattern of a rapid rise, a subsequent peak and an asymmetric contraction. This is
the central significance of the present work: despite the substantial variation in
historical context, chronology, geography and political organization, the fitted
trajectories reveal a common coarse-grained regularity in the rise and fall of
territorial systems.

The robustness of this pattern is particularly striking given the heterogeneity of
the examples considered. The maritime empires expanded through overseas control,
the land empires through contiguous conquest and frontier incorporation, and the
wartime cases through highly compressed episodes of military mobilization. Despite this, all these cases
can be represented within the same dynamical framework, with differences expressed
primarily through parameter values rather than through changes in functional form.
This suggests that the model captures structural features of expansion and
overextension that persist despite the many historical intricacies that distinguish
individual empires.

A second contribution of the paper is interpretive. By introducing \emph{chronophage}
as the mechanism governing asymmetric decline, the model provides a concrete
way of representing what has often been described more qualitatively as the rising
burden of complexity, coordination and maintenance. In the long-run historical
cases, this burden appears through administrative and fiscal expansion, whereas in
the wartime cases it is amplified by mobilization, occupation and the feedback of
international resistance. The same formalism therefore connects empirical fits to a
substantive account of why territorial growth eventually becomes self-limiting.

Since the model is analytically solvable, parameter-sparse and can be expressed in terms
of interpretable quantities, it allows for extension to additional historical and
contemporary cases. Its tractability makes it useful not only for retrospective
description but also for comparative analysis across systems that differ widely in size and duration.  The broader implication of these results is that the rise and fall of empires can be understood simultaneously as a statistical regularity and as a structural consequence of increasing complexity. The generalized Gompertz form identified in this study is not just an empirical fit: it also arises as a maximum-entropy, minimally biased envelope for asymmetric rise–peak–decline trajectories. At the same time, the chronophage gives this abstract form a concrete historical interpretation, by representing the cumulative burdens of administration, coordination and maintenance that grow with scale and ultimately outweigh the gains of further expansion.

In this way, our work places Tainter’s theory of collapse on a quantitative footing: diminishing returns on complexity are expressed dynamically as the gradual accumulation of internal constraints that first suppress expansion and then drive contraction. The common trajectory shared by otherwise very different empires therefore points to a general principle whereby large-scale social systems decline not only because of contingent shocks, but because complexity itself can become self-limiting. Thus, the minimal nature of the framework developed in this work, along with its wide applicability, makes it a valuable tool in understanding the evolution of human systems at large. 

\section*{Funding}
This publication is supported by the European Union's Horizon Europe research and innovation programme under the Marie Sk\l{}odowska-Curie Postdoctoral Fellowship Programme, SMASH co-funded under the grant agreement No. 101081355. The operation (SMASH project) is co-funded by the Republic of Slovenia and the
European Union from the European Regional Development Fund.

\section*{Disclaimer}
Co-funded by the European Union. Views and opinions expressed are however those of the author(s) only and do not necessarily reflect those of the European Union or European Research Executive Agency. Neither the European Union nor the granting authority can be held responsible for them.

\section*{Conflict of interest}
The authors declare no competing interests.

\section*{Data availability}

No new data were collected for this study. The code and processed data used to preprocess the historical series, fit the model, generate the figures, and reproduce the optimal-control simulations are available on Zenodo: \url{https://doi.org/10.5281/zenodo.22160644}

The code repository contains two main notebooks. One notebook reads in the pre-processed historical territorial time series, fits the closed-form rise--peak--decline model to the normalized data for each case, and reproduces the empirical figures reported in the paper. The second notebook implements the full optimal-control formulation, numerically integrates the coupled state--costate system, and generates the simulated trajectories and diagnostic outputs shown in the manuscript. Together, the code reproduces the model fitting, parameter estimation, figure generation, and optimal-control simulation underlying the results presented here.

\appendix

\section{Maximum-entropy characterization of the rise--fall envelope}
\label{app:maxent}

The functional form of Eq.~\eqref{eq:solution} can also be obtained from a simple maximum-entropy
construction \citep{KapurKesavan1992}. Let $x(t)\ge 0$ denote a non-negative temporal envelope, and define
its normalized profile $p(t)$ as:
\begin{equation}
p(t)=\frac{x(t)}{\int_{-\infty}^{\infty} x(u)\,du},
\qquad
\int_{-\infty}^{\infty} p(t)\,dt = 1.
\end{equation}
We seek the least-biased profile consistent with two weak structural
constraints: a fixed mean time and a fixed exponential moment. The mean fixes a
characteristic temporal location, while the exponential moment captures the
presence of an exponentially growing burden, hazard, or coordination cost. We
therefore maximize the Shannon entropy, $H[p]$:
\begin{equation}
H[p] = -\int_{-\infty}^{\infty} p(t)\ln p(t)\,dt
\end{equation}
subject to
\begin{equation}
\int p(t)\,dt = 1,
\qquad
\int t\,p(t)\,dt = \mu,
\qquad
\int e^{\alpha t} p(t)\,dt = m_\alpha,
\end{equation}
where $\alpha>0$ is treated as a fixed parameter, $\mu$ is the mean, and $m_\alpha$ is the prescribed
exponential moment. Introducing Lagrange multipliers $\lambda_0,\lambda_1,\lambda_2$, we extremize the resulting Lagrangian $\mathcal{L}[p]$:
\begin{equation*}
\mathcal{L}[p]
=
-\int p\ln p\,dt
-\lambda_0\!\left(\int p\,dt-1\right)
-\lambda_1\!\left(\int t\,p\,dt-\mu\right)
-\lambda_2\!\left(\int e^{\alpha t}p\,dt-m_\alpha\right).
\end{equation*}
Taking the first variation with respect to $p$ gives
\begin{equation}
\frac{\delta \mathcal{L}}{\delta p}
=
-\ln p(t)-1-\lambda_0-\lambda_1 t-\lambda_2 e^{\alpha t}=0.
\end{equation}
Hence
\begin{equation}
p(t)
=
\exp\!\left[-1-\lambda_0-\lambda_1 t-\lambda_2 e^{\alpha t}\right].
\end{equation}
For the empirically relevant unimodal case on $t\in\mathbb{R}$, we require
$\lambda_2>0$ and $\lambda_1<0$. Writing
\begin{equation}
\beta=-\lambda_1>0,
\qquad
\gamma=\lambda_2>0,
\end{equation}
$p(t)$ can now be rewritten as
\begin{equation}
p(t)=\frac{1}{Z}\exp\!\left[\beta t-\gamma e^{\alpha t}\right],
\end{equation}
where $Z$ is the normalization constant. Thus, the maximum-entropy profile is an exponential-of-exponential envelope. To connect this expression to the form used in the main text, note that the mode
$t_{\max}$ satisfies
\begin{equation}
\frac{d}{dt}\ln p(t)=\beta-\alpha\gamma e^{\alpha t}=0,
\end{equation}
so that
\begin{equation}
e^{\alpha t_{\max}}=\frac{\beta}{\alpha\gamma}.
\end{equation}
Now, define
\begin{equation}
s=\frac{\beta}{\alpha}>0,
\qquad
\gamma=s\,e^{-\alpha t_{\max}};
\end{equation}
this gives
\begin{equation}
p(t)\propto
\exp\!\left[s\alpha(t-t_{\max})-s\,e^{\alpha(t-t_{\max})}\right].
\end{equation}
Finally, the empirical trajectory $x(t)$ is an unnormalized envelope rather than
a probability density, so we restore an arbitrary vertical scale and impose the
peak condition $x(t_{\max})=x_{\max}$. This yields
\begin{equation}
x(t)
=
x_{\max}
\exp\!\left[
s+s\alpha(t-t_{\max})-s\,e^{\alpha(t-t_{\max})}
\right],
\end{equation}
which is exactly the functional form used in Eq.~\eqref{eq:solution}.

In this sense, Eq.~\eqref{eq:solution} can be interpreted as the
minimum-bias, or maximum-entropy, unimodal asymmetric envelope compatible with
a fixed temporal location and a fixed exponential moment. The resulting form
does not impose any additional structure beyond these constraints, and therefore
provides an information-theoretically neutral baseline for the observed
rise--peak--decline trajectories.

\section{Optimal control formulation and reduced dynamics}
\label{app:control}

Following standard textbook treatments of infinite-horizon optimal control, we
formulate territorial expansion as an optimal control problem in which
territorial extent $x(t)>0$ is supported by territorial support capacity $y(t)>0$
\citep{KamienSchwartz1991,SeierstadSydsaeter1987,SethiThompson2000}.
Both $x(t)$ and $y(t)$ are measured in units of area. Territory grows at a
saturating rate relative to available capacity, while capacity itself can be
adjusted through policy effort. The state equations are
\begin{equation}
\dot{x}(t)=r\,x(t)\left(1-\frac{x(t)}{y(t)}\right),
\qquad
\dot{y}(t)=v(t)\,y(t),
\label{eq:oc_states}
\end{equation}
where $r>0$ is the intrinsic expansion rate and $v(t)$ is the proportional
adjustment rate of support capacity.

The empire chooses the control $v(t)$ to maximize the discounted lifetime
objective \citep{KamienSchwartz1991}:
\begin{equation}
J=\int_0^\infty e^{-\rho t}L(x(t),y(t),v(t))\,dt.
\label{eq:oc_objective}
\end{equation}
Here $J$ denotes the total discounted net payoff from managing territorial
extent and support capacity over time, $\rho>0$ is the discount rate, and
$L(x(t),y(t),v(t))$ is the instantaneous or running payoff. Maximizing $J$
therefore means choosing a policy that evolves over time, so as to balance the gains
from territorial expansion against the costs of sustaining and adjusting the
capacity required to support that expansion.

We propose the following ansatz for the running payoff, $L(x(t),y(t),v(t))$, in Eq.~\eqref{eq:oc_objective}:
\begin{equation}
L(x,y,v)
=
C\frac{x^2}{y}
+
\kappa\left[\left(r-\gamma\right)\left(1-\frac{x}{y}\right)-\alpha\right]v
-
\frac{\kappa}{2}v^2,
\label{eq:oc_payoff}
\end{equation}
with $C,\kappa>0$, $\gamma\ge0$, and $\alpha\ge0$. The parameter $C$ scales
the direct return to effective territorial control, which is increasing in
territorial extent but moderated by the support capacity available to sustain
it. The parameter $\kappa$ sets the overall strength of capacity adjustment,
including both the incentive to change support capacity and the quadratic cost
that makes rapid adjustment increasingly expensive. The parameter $\gamma$
governs the strength of the balancing force that tends to restore proportionality
between territorial extent and support capacity, while $\alpha$ represents the
increasing burden associated with sustaining territorial overextension.

To characterize the optimal policy, it is useful to introduce the Hamiltonian,
which combines the current payoff with the effect of present choices on the
future evolution of the state variables \citep{KamienSchwartz1991,SethiThompson2000}.
Using current-value costates $\mu_x$
and $\mu_y$, the Hamiltonian is
\begin{equation}
H(x,y,v,\mu_x,\mu_y)
=
L(x,y,v)
+
\mu_x r x\left(1-\frac{x}{y}\right)
+
\mu_y v y .
\label{eq:oc_hamiltonian}
\end{equation}
The costates (also referred to as shadow values in the literature) $\mu_x$ and $\mu_y$
measure the marginal value of an additional unit of
territorial extent and support capacity, respectively \citep{KamienSchwartz1991,SeierstadSydsaeter1987}.

Pontryagin's maximum principle \citep{SethiThompson2000}
provides the necessary conditions that any
optimal path must satisfy: the state equations, the costate equations, and the
stationarity condition for the control. These are
\begin{equation}
\dot{x}=\frac{\partial H}{\partial \mu_x}, \qquad
\dot{y}=\frac{\partial H}{\partial \mu_y}, \qquad
\dot{\mu}_x=\rho\mu_x-\frac{\partial H}{\partial x}, \qquad
\dot{\mu}_y=\rho\mu_y-\frac{\partial H}{\partial y}, \qquad
\frac{\partial H}{\partial v}=0.
\label{eq:oc_pontryagin}
\end{equation}
The stationarity condition in Eq.~\eqref{eq:oc_pontryagin} gives the following optimal feedback rule:
\begin{equation}
v^\ast
=
\left(r-\gamma\right)\left(1-\frac{x}{y}\right)-\alpha
+
\frac{\mu_y y}{\kappa}.
\label{eq:oc_feedback}
\end{equation}
The last term, $\mu_y y/\kappa$, reflects the forward-looking value of expanding support capacity:
when additional capacity is valuable in future periods, the optimal policy
raises $v^\ast$ accordingly. Since $\mu_y$ is the costate associated with $y$, it measures the
marginal contribution of an additional unit of support capacity to the
discounted objective. Thus, $\mu_y y/\kappa$ captures the extent to which the
empire increases current adjustment effort because extra capacity is valuable
for future expansion. To isolate the core rise--fall mechanism, it is useful to work with the
congestion ratio
\begin{equation}
z(t)=\frac{x(t)}{y(t)} .
\label{eq:z_def}
\end{equation}
This variable measures the degree to which territorial extent outpaces support
capacity. Rewriting the dynamics in terms of $z(t)$ clarifies that the
central mechanism is the accumulation of mismatch between expansion and the
capacity required to sustain it. Using Eq.~\eqref{eq:z_def} in Eq.~\eqref{eq:oc_states}, we obtain
\begin{equation}
\dot{z}
=
z\left[r(1-z)-v^\ast\right].
\label{eq:z_general}
\end{equation}
Substituting Eq.~\eqref{eq:oc_feedback} into
Eq.~\eqref{eq:z_general} gives
\begin{equation}
\dot{z}
=
\gamma z(1-z)+\alpha z-\frac{\mu_y y}{\kappa}z.
\label{eq:z_exact}
\end{equation}
To obtain a transparent reduced form, we consider the
regime in which the term $\mu_y y/\kappa$ is small relative to the direct
state-dependent terms during the initial transient. This approximation isolates
the basic mechanism by which mismatch between territorial extent and support
capacity evolves. Eq.~\eqref{eq:z_exact} then reduces to
\begin{equation}
\dot{z}
=
\gamma z(1-z)+\alpha z .
\label{eq:z_reduced}
\end{equation}

Eq.~\eqref{eq:z_reduced} contains two distinct mechanisms. The term
$\gamma z(1-z)$ represents a logistic contribution, which tends to restore
balance between territorial extent and support capacity. The term
$\alpha z$ captures the self-reinforcing growth of congestion, that is, the
accumulation of mismatch between territorial integration and administrative
capability.

\begin{figure}[t]
    \centering
    \includegraphics[width=\linewidth]{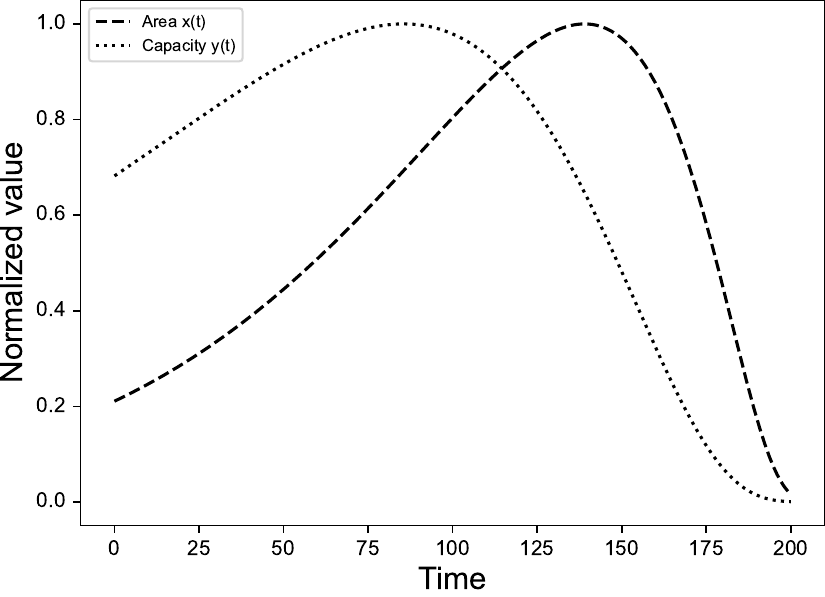}
    \caption{\textbf{Simulated trajectories of territorial extent and support capacity.}
    The figure displays the full optimal-control dynamics for territorial extent $x(t)$ (dashed) and support capacity $y(t)$ (dotted). Both $x(t)$ and $y(t)$ are normalized, whereas time is expressed in number of years, starting at zero. Territorial extent grows at a saturating rate relative to available support, whereas support capacity evolves through the control variable. Both trajectories are normalized to a maximum of one to emphasize their relative temporal profiles.}
    \label{fig:xy_optimal_control}
\end{figure}

\paragraph{Logistic case ($\alpha=0$).}

When the mismatch term vanishes ($\alpha=0$), Eq.~\eqref{eq:z_reduced} reduces to
\begin{equation}
\dot{z}=\gamma z(1-z);
\label{eq:z_logistic}
\end{equation}
this is the standard logistic equation. Its solution is
\begin{equation}
z(t)=
\frac{1}{1+A e^{-\gamma(t-t_0)}},
\qquad
A=\frac{1-z_0}{z_0}.
\label{eq:z_logistic_solution}
\end{equation}
Using
\begin{equation}
\frac{\dot{x}}{x}=r(1-z),
\label{eq:x_growth}
\end{equation}
integration yields
\begin{equation}
x(t)
=
x_0
\left[
\frac{1+A}{1+A e^{-\gamma(t-t_0)}}
\right]^{r/\gamma}.
\label{eq:x_logistic_solution}
\end{equation}
Thus, when $\alpha=0$, the model produces a symmetric logistic-type
saturation trajectory for territorial expansion.

\paragraph{Rise--fall case ($\alpha>0$).}

When mismatch accumulates ($\alpha>0$), the additional term in
Eq.~\eqref{eq:z_reduced} contributes to the growth of the congestion ratio, $z(t)$.
In the limiting regime where this term dominates the restoring mechanism,
Eq.~\eqref{eq:z_reduced} reduces to
\begin{equation}
\dot{z}=\alpha z,
\label{eq:z_risefall}
\end{equation}
with solution
\begin{equation}
z(t)=z_0 e^{\alpha(t-t_0)} .
\label{eq:z_risefall_solution}
\end{equation}
Substituting Eq.~\eqref{eq:z_risefall_solution} into Eq.~\eqref{eq:x_growth} gives
\begin{equation}
\dot{x}
=
r x\left(1-z_0 e^{\alpha(t-t_0)}\right).
\end{equation}
Integrating this gives us
\begin{equation}
x(t)
=
x_0
\exp\left[
r(t-t_0)
-
\frac{r z_0}{\alpha}
\left(e^{\alpha(t-t_0)}-1\right)
\right].
\label{eq:x_risefall_raw}
\end{equation}
Let $r=s\alpha$; recentering time at the peak
$t_{\max}$ defined by $z(t_{\max})=1$ yields
\begin{equation}
x(t)
=
x_{\max}
\exp\left[
s+s\alpha(t-t_{\max})
-
s e^{\alpha(t-t_{\max})}
\right],
\label{eq:x_risefall_final}
\end{equation}
which is the functional form used in Eq.~\eqref{eq:solution}.

The two behaviors therefore arise within the same dynamical framework.
When $\alpha=0$, territorial expansion approaches a symmetric logistic
saturation. When $\alpha>0$, the accumulation of mismatch between
territorial reach and support capacity drives an asymmetric
rise--peak--decline trajectory.

\bibliographystyle{unsrt} 
\bibliography{biblio}

\end{document}